\documentclass[12pt]{article}
\usepackage{amsfonts}
\usepackage{amssymb}
\usepackage{epsfig}
\usepackage{delarray}
\newcommand{\beq}{\begin{equation}}
\newcommand{\eeq}{\end{equation}}

\def\simle{\mathrel{
   \rlap{\raise 0.511ex \hbox{$<$}}{\lower 0.511ex \hbox{$\sim$}}}}

\newcommand{\al}{\alpha}
\newcommand{\be}{\beta}

\newcommand{\ga}{\gamma}

\newcommand{\De}{\Delta}
\newcommand{\de}{\delta}

\renewcommand{\epsilon}{\varepsilon}

\newcommand{\MeV}{\mbox{\rm MeV}}
\newcommand{\s}{\mbox{\rm s}}
\newcommand{\keV}{\mbox{\rm keV}}
\newcommand{\cm}{\mbox{\rm cm}}
\newcommand{\GeV}{\mbox{\rm GeV}}

\begin{document}

\begin{titlepage}
\vspace*{\fill}
\begin{center}
\LARGE Time Delay of Radiation from Gamma-Ray Burst Sources as a
Test for a Model of the Universe
\end{center}
\vspace{0.9\fill}
\begin{center}
\large Yu.S. Grishkan, \quad V.B. Petkov${}^{*}$, \protect\\
E.G. Vertogradova
\end{center}
\vspace{-5mm}
\begin{center}
\large Institute for Nuclear Research of RAS
\end{center}
\vspace{\fill} The estimation of light velocity dispertion caused
by quantum fluctuations influence on electromagnetic wave
propagation in four-dimensional space-time is presented.
Analytical cosmological solutions for the flat and open Universe
for the case of cosmic vacuum model are obtained. The time delay
of cosmic gamma-ray burst radiation is calculated. The delay is
explained by propagation of dispersive electromagnetic wave in the
expanding Universe. It is shown that the delay value depends on a
model of the Universe. We conclude that to discern a model of the
Universe measurement accuracy of parameter $\De t/\De E_{\ga}$
should be better than $10^{-5} \; \s / \MeV$. \vspace{\fill}
\begin{flushleft}
\end{flushleft}
\vspace{\fill}
\begin{flushleft}
${}^{*}$\small{E-mail: petkov@neutr.novoch.ru}
\end{flushleft}
\end{titlepage}

\section{Introduction}
Recently a hypothesis was suggested that global Lorentz invariance
is only an approximate symmetry of nature and may be broken for
elementary particles participating in various physical
interactions in case of high enough energy of the particle
(Amelino-Camelia et al 1998, Coleman and Glashow 1999, Stecker and
Glashow 2001). As the process of this type we consider a mechanism
of generation of Lorentz-noninvariant space-time metrics
components caused by quantum fluctuations (Ellis et al 1998). The
microscopic quantum fluctuations, which may occur on scale sizes
of order the Planck length $ l_{pl} \sim {10}^{-33} \cm $, have
affected fundamentally the large-scale space-time structure at the
early stages of the universe creation (Hawking et al) and have
caused, in particular, the universe birth from vacuum and the
stage of quantum inflation (Vertogradova and Grishkan 2000). But
now the same fluctuations also exists and may have an influence on
the rays of light propagating through the space-time. Indeed as it
is shown by Ellis et al (1998, 2000a) the fluctuations,
constituting the so-called space-time foam, transform macroscopic
properties of space-time metrics and generate non-diagonal
components of the metrics $g_{0 \al}$, so that
$$
g_{0 0}=-1, \quad g_{\al \be}={\de}_{\al \be}, \quad g_{0
\al}=\frac{u_{\al}}{c}
$$
Here Greek letters have values $\al, \be = 1,2,3$, $c$ is the
velocity of light constant. As a result, the whole space-time is
rotated at a velocity proportional to $u_{\al} / c \ll 1$. A
vector of rotation velocity $u_{\al}$ breaks global Lorentz
invariance (Coleman and Glashow 1999) and thus transforms
dispersion properties of light waves propagating through the
space-time
\beq
c(E)=c \left( 1- \frac{u}{c} \right)
\eeq
where $E = \hbar \omega$ is the photon energy of frequency
$\omega$, $u$ is the absolute value of the vector $u_{\al}$. The
quantum-gravity theory, particularly the space-time foam theory
(Ellis et al 2000a), predicts the ratio of velocities appearing in
(1) is proportional to the ratio of the photon energy $E$ to
effective quantum-gravity energy scale $M_{QG}$
\beq
\frac{u}{c} \sim \frac{E}{M_{QG}}
\eeq
Consequently the delay in the arrival times of photons of
different energies and hence velocities propogating in the
space-time foam from distant source of radiation is estimated as
\beq
\De t = \frac{\De L (\De u)}{c} \sim \frac{L}{c}
\frac{\De u}{c} \sim \frac{L}{c} \frac{\De E_{\ga}}{M_{QG}}
\eeq
where $L$ is a source distance, $\De E_{\ga}$ is the difference of
energies of detected photons.

As the radiation from distant sources propagates through the
space-time foam there must be the delay in the arrival times of
photons depending on the photon energy. If we succeed in measuring
this delay we will be able to test quantum-gravity ideas. Moreover
it is the constant $M_{QG}$ that characterize the energy scale of
the effects of quantum gravity. Therefore it will be possible to
estimate the quantum-gravity energy scale $M_{QG}$ in the
following way
\beq
M_{QG} = \frac{L}{c} \frac{\De E_{\ga}}{\De t}
\eeq
It is worth noting that the ratio $E / M_{QG}$ is negligible. The
time delay effect may be observable only for cosmological
distances. It is also clear that the higher the energy of
radiation the better. Therefore $\ga$ -- ray bursts sources of
extragalactic origin are considered as high-energy radiation
sources of interest. The time delay in the arrival times of
photons of different energies from $\ga$ -- ray burst sources was
studied by Ellis et al (2000b) based on BATSE and OSSE
observations (Paciesas et al 1999, OSSE Collaboration 1999). The
calculation of the source distance in formulas (3), (4) may be
realized for certain model of the universe. In this way
theoretical calculation of the time delay is universe-model
dependent. The approach described above implies that the photons
of different energies are radiated by the source of $\ga$ -- ray
burst at the same time and the photons time delay is the effect of
quantum gravity and is not caused by processes inside the source
of radiation.

\section{Time~delay-redshift diagram -- theoretical calculation}
We obtain here the time delay value of radiation in the expanding
universe in terms of the cosmic vacuum model (Chernin 2001). The
exact first integral of Einstein's equation for this model is
\beq
\frac12 \; {\dot{a}}^2 = \frac12 \; A_V^{-2} a^2 + (A_D+A_B) \; a^{-1} +
\frac12 \; A_R^2 a^{-2} - \frac12 \; k
\eeq
where $a(t)$ is the scale factor of the universe, $k=0,\pm 1$ is
the space curvature sign (it corresponds to flat, closed and open
universe). The derivative in equation (5) means $\dot{a} = da /
d(ct)$, where $t$ is physical time. The values of the Friedmann's
integrals are obtained experimentally: $A_V \sim a_0 \sim
{10}^{28} \; \cm $ is the cosmic vacuum constant, $A_D \sim
{10}^{-1} A_V $ is the dark matter constant, $A_B \sim {10}^{-2}
A_V $ is the constant of baryon's subsystem, $A_R \sim {10}^{-2}
A_V $ is the constant of radiation of all types, $a_0$ is the
current value of the scale factor of the universe (which is
calculated for flat model of Friedmann type $a_0 \simeq t_0 \, c$)
and the current age of the universe is $t_0 \sim 2.8 \times
{10}^{17} \s \sim 10 \, \mbox{Gyr} $. The model of Friedmann type
means here matter-dominated epoch of the universe evolution.

Within the framework of model under discussion the value of the
space curvature sign $k$ is usually ignored because the sign have
not significant influence on the fate of the cosmic expansion and
hence on the dynamics of observable cosmological objects. We will
show however that the influence is very important for the precise
effect of time delay. Therefore the $k$ value will not be
neglected below. We consider here the values $k=0$ (flat model of
cosmic vacuum) and $k=-1$ (open model of cosmic vacuum). As
regards the closed model $k=1$, theoretical interpretation of time
delay effect for this model is difficult for many reasons.

We consider baryon-vacuum epoch of the universe evolution. The
scale factor during the epoch of interest is $a \gg A_R^2 / A_D$
($z \ll 1000$), so the constant of radiation $A_R$ in equation (5)
is appropriate to be neglected. Therefore we can find the time $t$
for given value of the scale factor $a$ in the form
\par\medskip\noindent
\beq
\frac{t-t_0}{t_0} = \int_{a_0}^{a}
\frac{ d (a/a_0) }{ \left [ \left( a / a_0 \right )^2 +
\left( a_1 / a_0 \right) \left( a / a_0 \right )^{-1} - k \right
]^{1/2} }
\eeq
\par\medskip\noindent
where $a_1 = 2 \; (A_D+A_B)$. Besides the analytical solution in
terms of primitive functions is obtained for $k=0$
\beq
\frac{t-t_0}{t_0} = \frac23 \ln
\left \{ \frac{ \left ( a / a_0 \right )^{3/2} + \left [ \left ( a
/ a_0 \right )^3 + \left( a_1 / a_0 \right) \right ]^{1/2} }{1+
\left [ 1+ \left( a_1 / a_0 \right) \right ]^{1/2} } \right \}
\eeq
The scale factor dependence in (6), (7) can be parametrized by the
redshift $z$ according to standard formula $ a / a_0 = 1/ (1+z) $.
As a result the integral in equation (6), for example, takes the
form
\beq
\frac{t-t_0}{t_0} = - \int_1^{z+1} \frac{d \xi}{\xi
\left[ 1 - k {\xi}^2 + \left( a_1 / a_0 \right) {\xi}^3
\right]^{1/2} }
\eeq

Taking into account expressions (1)--(3) and (8) we obtain the
time delay of radiation from distant source $\De t$ in the form
\par\bigskip\noindent
$$
\frac{\De t}{t_0} = \int_1^{z+1} \frac{\De u}{c} \frac{d \xi}{\xi
\left[ 1 - k {\xi}^2 + \left( a_1 / a_0 \right) {\xi}^3
\right]^{1/2} }
$$
or
\beq
\frac{\De t}{t_0} = \frac{\De E_{\ga}}{M_{QG}} \; \Phi(z,k)
\eeq
\par\medskip\noindent
$$
\mbox{where} \quad \Phi(z,k) = \int_1^{z+1} \frac{d \xi}{ \left[ 1
- k {\xi}^2
+ \left( a_1 / a_0 \right) {\xi}^3 \right]^{1/2} }
$$
\par\bigskip\noindent
Here we have taken into consideration the fact that the photon
energy varies during the process of the Universe evolution
depending on the redshift as $E=E_{\ga} (1+z)$.

Up to a constant the asymptotic form of time delay formula (9) for
high-redshift objects $z \rightarrow \infty$ is in agreement with
appropriate expression obtained by Ellis et al (2000b).
\par\bigskip\noindent
\beq
\frac{\De t}{t_0} = \frac{\De E_{\ga}}{M_{QG}} \; z, \quad z \ll 0.7 \quad (z \rightarrow 0)
\eeq
\beq
\frac{\De t}{t_0} = \frac{\De E_{\ga}}{M_{QG}} \left(
\displaystyle\frac{a_0}{a_1} \right)^{1/2} 2 \; \tilde{z}, \quad z
\gg 0.7 \quad (z \rightarrow \infty)
\eeq
\par\bigskip\noindent
where $\tilde{z}=1-(1+z)^{-1/2}$ and $a_0/a_1 \sim 5$. The formula
(10) corresponds to the universe expansion according to the law of
Hubble and formula (11) -- to the universe expansion of flat
Friedmann type.

We'd like to note that the model of cosmic vacuum gives the
expression (11) for the time delay only at the limit of high
redshift $z\rightarrow \infty$. Because of this the time
delay-redshift dependence is different for the cosmic vacuum model
and for the flat Friedmann model discussed by Ellis et al (2000b).
This fact is quite important for experimental data analysis.

\section{The time delay experimental data and the possibility of determination of the fate of the cosmic expansion}
The Picture 1 illustrates theoretical dependences of parameter
$\De t / \De E_{\ga}$ as a function of redshift $z$. The
dependences were computed from formula (9) for different
cosmological models: Curve 1 corresponds to the flat model of
Friedmann type, Curve 2 corresponds to the flat model of cosmic
vacuum and Curve 3 corresponds to the open model of cosmic vacuum.
The effective quantum-gravity energy scale is set to be equal to
the Planck energy scale $10^{19} \; \GeV$. From this Picture it is
possible to determine the fate of the cosmic expansion when the
measurement accuracy of parameter $\De t / \De E_{\ga}$ is better
than $10^{-5} \; \s / \MeV$.

Thus for the cosmological model selection (or for estimation of
the quantum-gravity energy scale) measurements of $\ga$
-- ray bursts redshifts and the comparative time delay of photons
of different energies are required. The paper of Ellis et al.
(2000b) contains experimental data from BATSE catalog (Paciesas et
al. 1999) and OSSE data (OSSE Collaboration 1999) for five $\ga$
-- ray bursts whose redshifts are known:\quad 1)~$\rm GRB970508$,
$z=0.835$,\quad 2)~$\rm GRB971214$, $z=3.14$,\quad 3)~$\rm
GRB980329$, $z=5.0$\quad 4)~$\rm GRB980703$, $z=0.966$,\quad
5)~$\rm GRB990123$, $z=1.60$.\quad The energy ranges of photons
from this $\ga$ -- ray burst sources observed by BATSE are:
Channel 1 between $20$ and $50 \; \keV$ and Channel 3 between
$100$ and $300 \; \keV$. The difference of arrival times for
Channel 1 and Channel 3 photons is of our interest. Moreover two
$\ga$
-- ray bursts -- $\rm GRB980329$ and $\rm GRB990123$ -- were also
detected by OSSE detector (OSSE Collaboration 1999) in a single
channel with energy range $1-5 \; \MeV$. The experimental time
delay in this case is the difference of arrival times of photons
for OSSE data and Channel 3 of BATSE. Both BATSE and OSSE
experimental values of parameter $\De t / \De E_{\ga}$ as a
function of redshift $z$ obtained according to the paper of Ellis
et al. (2000b) are given at the Picture 2. We compare Picture 1
with Picture 2 and conclude that at present moment measurement
accuracy does not allow to determine the fate of the universe
expantion.

\section{Conclusions and prospects}
We believe that analysis of more statistically significant number
of $\ga$ -- ray bursts in the future with measured redshifts and
delays in the arrival times of photons of different energies will
allow us to realize theory-experiment comparison more effectively.
When this reliable data appear it will be possiable to test
quantum-gravity effects, select cosmological models of different
topology of embedded three-dimensional space ($k=0,-1,1$) and
obtain more precise values of the approximatly known Friedmann's
integrals, particularly the integral $A_V$. For the time being we
can only use the available experimental data (Ellis et al. 2000b)
and estimate the quantum-gravity energy scale as
$$
M_{QG} \gtrsim  {10}^{15} \; \GeV
$$

In the case of $ M_{QG} \sim M_{pl} \sim {10}^{19} \; \GeV $ we
conclude that for the investigation of effects of delays in the
arrival times of photons from cosmological $\ga$ -- ray burst
sources measurement accuracy of ratio $\De t / \De E_{\ga}$ should
be better than $\sim 10^{-5} \; \s / \MeV$.

The research was supported by grant RFBR 00-02-16095.

\section*{References}

\begin{trivlist}
\item[] {\it Amelino-Camelia~G., Ellis~J., Mavromatos~N.E., Nanopoulos~D.V.,
Sarkar~S.)} // Nature, 1998, v.~393, p.~763.
\item[] {\it Chernin A.D.} // Uspehi phis. nauk, 2001, v.~171, N11, p.~1153.
\item[] {\it Coleman S., Glashow S.L.} // Phys. Rev. D, 1999, v.~59, p.~116008.
\item[] {\it Ellis J., Kant P., Mavromatos~N.E., Nanopoulos~D.V.,
Winstanley~E.} // Mod. Phys. Lett., 1998, v.~A13, p.~303.
\item[] {\it a Ellis J., Mavromatos~N.E., Nanopoulos~D.V.}
// CERN-2000-136; preprint gr-qc/0005100.
\item[] {\it b Ellis J., Farakos~K., Mavromatos~N.E., Mitsou~V.A.,
Nanopoulos~D.V.} // Astrophys. J., 2000, v.~535, p.~139.
\item[] {\it Hawking S., Page D.N., Pope C.N.} // Nucl. Phys.
B, 1980, v.~170 [FSH], p.~283.
\item[] {\it OSSE Collaboration} // Gamma Ray bursts time profiles, 1999;
\linebreak http://www.astro.nwu.edu/astro/OSSE/bursts/.
\item[] {\it Paciesas W.S. et al} // Astrophys. J. Suppl. Ser., 1999,
v.~122, p.~465.
\item[] {\it Stecker F.W. and Glashow S.L.} // Astropart. Phys., 2001,
v.~16, p.~97.
\item[] {\it Vertogradova E.G., Grishkan Yu.S.} // Astronomy Reports, 2000,
v.~77,
N3, p.~1.
\end{trivlist}

\begin{figure}[htb]
\epsfxsize=\textwidth \epsfbox{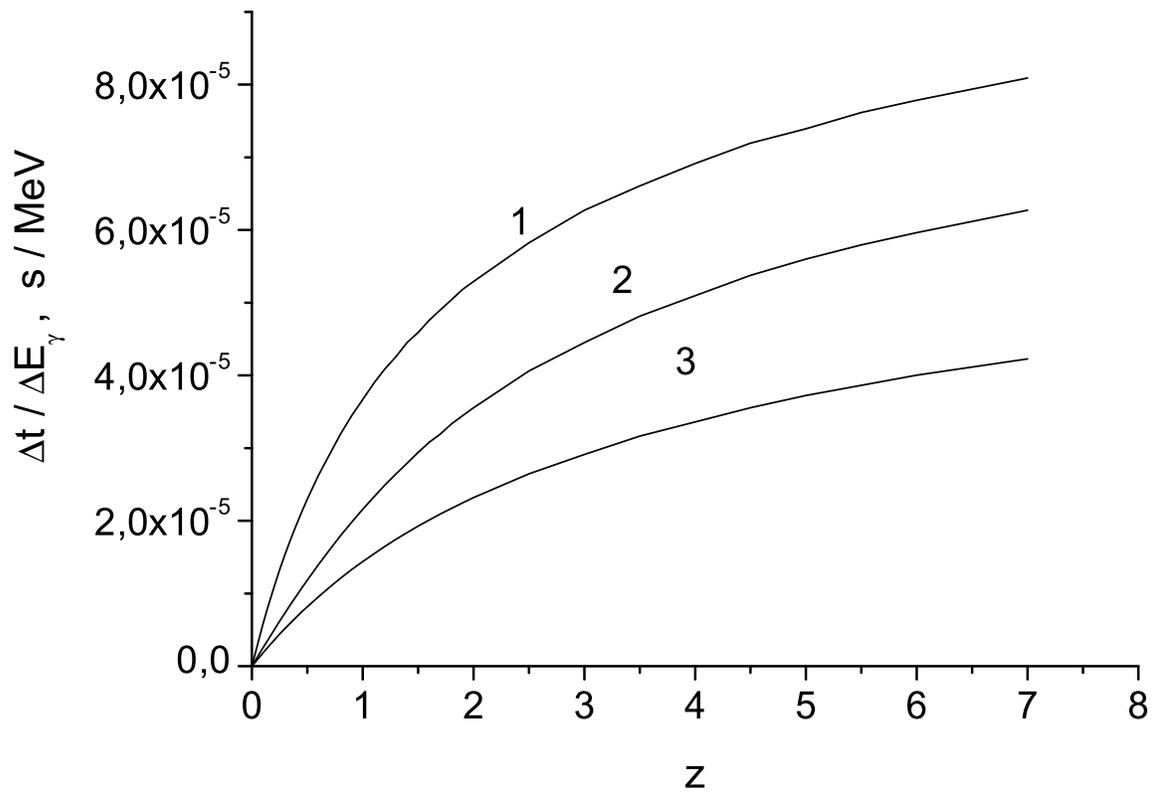} \caption{Theoretical
Results. Line 1 corresponds to the solution of Flat Friedmann
type, Line 2 - to the Flat Model of Cosmic Vacuum and Line 3 - to
the Open Model of Cosmic Vacuum}
\end{figure}

\begin{figure}[htb]
\epsfxsize=\textwidth \epsfbox{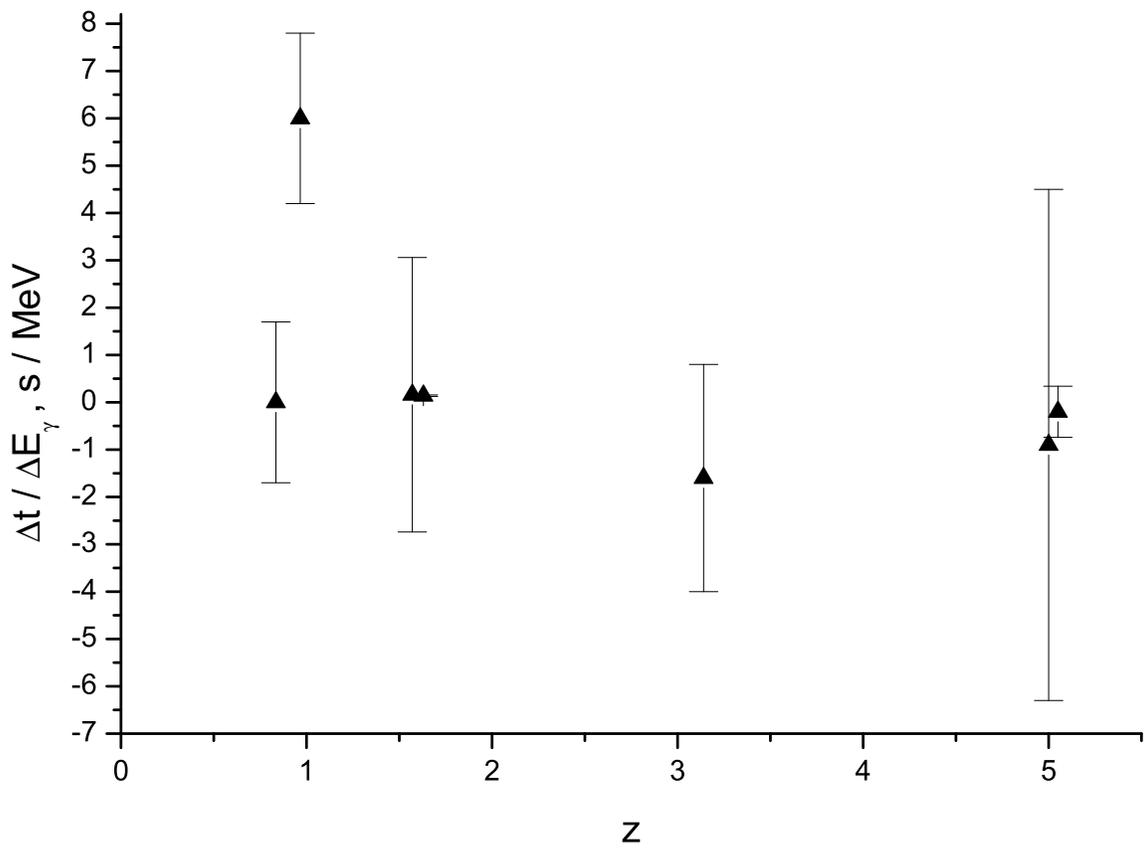} \caption{Results of
Fits to the Gamma-Ray Burst Data from BATSE and OSSE}
\end{figure}

\end{document}